# Beyond Bias: Participatory and Reflective Approaches to Cultural AI

Archana Prasad
Royal College of Art
London, UK
archana.prasad@network.rca.ac.uk

Isha Singh
Gooey.AI
Lucknow, India
coordinator@dara.network

Tom Simmons
Royal College of Art
London, UK
tom.simmons@rca.ac.uk

## ABSTRACT

Generative AI systems increasingly shape cultural production, yet creative intentions, cultural meanings, and interpretive practices often can't be articulated through computational metrics alone. This paper presents Beyond Bias, a collaboration between Gooey.AI and Goethe-Institut India, as a participatory approach to cultural AI which includes collaborative dataset creation, reflective AI tooling, artist-led model fine-tuning, and co-authored governance practices. Across 9 workshops involving over 200 participants, artists and cultural practitioners engaged with AI systems through experimentation,iteration, and collaborative LoRA training. Participants used their AI-generated outputs and visualizations as reflective interfaces for exploring symbolism, memory, authorship, and cultural contexts. Comparing contemporary generative AI outputs with participant fine-tuned outputs helped participants reflect on cultural details missing in big tech AI systems.

This paper contributes reflective AI tooling approaches foregrounding transparency, stewardship, and community participation; findings from participatory workshops examining how generative AI visualizations mediate cultural representation and interpretive practice; and a framework for cultural AI grounded in cultural integrity, and reflective practice.

## KEYWORDS

Cultural AI, Human-AI Co-Creation, Reflective Creative Practice, Participatory Design, Creative Agency, Interpretive AI, Creative Interaction, AI and Creativity, Community-Centered AI, Generative AI

## 1 Introduction

Generative AI systems increasingly participate in cultural production. Image, language, and video models shape aesthetic norms, social narratives, and forms of expression across everyday life. As these systems become embedded within creative practice, they must be understood not only as computational tools but also as cultural technologies that mediate representation and meaning-making [1,4].

Current approaches to evaluating generative AI systems have primarily focused on harm mitigation, including bias, toxicity, and misinformation. While these approaches remain necessary, they imply a limited definition of cultural success: systems are considered successful when they avoid harmful outputs [2]. This framing leaves a broader question unanswered: what does it mean for AI systems to engage culture well?

Generative models frequently reproduce dominant Western visual and linguistic norms while flattening regional aesthetics and contextual forms of knowledge into generalized representations [1,3]. Cultural forms are often reduced to stylistic markers detached from lived practices, histories, and symbolism or erased altogether.[4]

This paper presents Beyond Bias, a collaborative initiative between Gooey.AI and Goethe-Institut India launched in March 2025, exploring participatory approaches to cultural AI through collaborative dataset creation, reflective tooling, and artist-led fine-tuning.

## 2 Cultural and Participatory Approaches

Research on fairness and bias in AI has demonstrated how machine learning systems reproduce inequalities embedded within datasets and computational infrastructures [2]. Generative systems have been shown to privilege dominant Western visual and linguistic norms while not engaging with under-represented communities. Image generation by big tech models often shows stereotyped outputs for certain communities. Eg. Indian women may be represented as clad in sarees, with bindis and luscious black hair.

Recent scholarships increasingly frame generative AI systems as cultural technologies rather than purely informational systems. Crawford describes AI infrastructures as deeply entangled with social, political, and material systems [1]. Critical scholars have also argued that contemporary AI infrastructures frequently reproduce extractive relationships between platforms, datasets, and communities [3,6]. These concerns become particularly important in cultural contexts where local knowledge, aesthetics, and histories may be absorbed into AI datasets without meaningful consent, attribution, or governance.

Participatory and human-centered approaches to AI development emphasize stakeholder involvement, situated expertise, and collaborative design [5]. Through Beyond Bias, we build upon participatory approaches by positioning artists and cultural practitioners not only as users of AI systems, but as co-creators and active participants in dataset creation, fine-tuning, interpretation, and governance.

## 3 The Beyond Bias Initiative

Beyond Bias involved over 200 participants and 24 organizations across nine workshops involving collaborative dataset creation, prompt experimentation, LoRA training, and reflective discussion. We co-created an open manifesto with participants and used that as scaffolding to inform the development of AI tools.

The initiative began by creating a constoria of institutional stakeholders across US, UK, EU and India. Over two roundtables they collaboratively authored an open manifesto articulating principles for culturally grounded AI development for model makers. Themes included cultural integrity, transparency, environmental accountability, fair compensation, and community stewardship. The manifesto functioned not only as an ethical framework and instigation for technical development of AI tools for cultural practitioners but also as a governance experiment that foregrounded collective authorship and participatory decision-making within AI development processes. Each Beyond Bias workshop was an iteration on the previous one, and led to updates in our two major tools - The AI Image Trainer tool and the Video Generation tool.

### 3.1 Reflective Tool Design

A central component of the initiative involved the development of reflective AI tools designed not only for generation, but also for transparency, participation, and stewardship.
The project developed accessible image fine-tuning and video-generation tools that enabled participants without extensive tech expertise to train LoRAs using collaboratively curated datasets. Participants could experiment with multilingual prompting workflows, including Hindi-to-English prompt translation systems designed to support under-resourced languages. Images were trained using Flux, and participants could choose from multiple models to generate their outputs, thus supporting creative agency and intent.

The tools also incorporated low-resolution generation options and lightweight pathways that foregrounded environmental awareness. Tools shared the eco cost of each run, in terms of water and electricity usage. This reflected broader concerns around the material and ecological costs of AI infrastructures [1]. Discussions surrounding privacy, consent, and cultural extraction informed the sessions as well. Our recent Beyond Bias workshop at the Royal College of Art, conducted in collaboration with the Mozilla Foundation, enabled participants to choose whether their generated datasets could be shared publicly and contribute to the training of future AI models.

Rather than positioning participants solely as end-users of AI systems, the project explored how localized fine-tuning and participatory tooling could create pathways for communities to influence broader generative ecosystems.

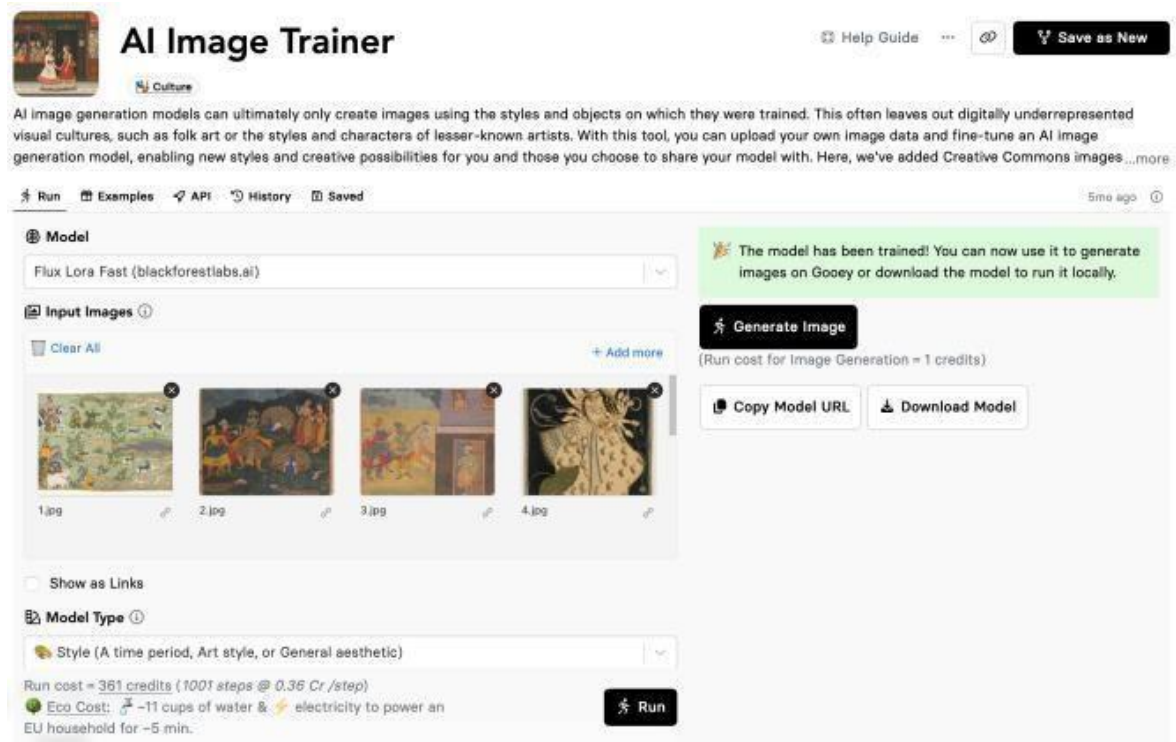

Figure 1: AI Image Trainer Tool with Ecological Costs Displayed at the Bottom

### 3.2 Participatory Workshops and Cultural Outputs

Participants frequently observed that LLMs produced flattened and generalized representations of culturally specific aesthetics. Prompts referencing Mughal miniatures, regional folk art, or embroidery traditions often generated outputs that appeared visually ornate while overlooking compositional structure, symbolism, material texture, and regional variation.

For example, prompts related to Mughal miniature traditions frequently resulted in generic “Indian royal court” imagery characterized by symmetrical palace settings, decorative clothing, and homogenized visual motifs. Participants noted that these outputs lacked the narrative density, and layered symbolism associated with miniature traditions. Outputs associated with “Indian art” often emphasized saturated colors, exoticized ornamentation, or spiritual symbolism. Similarly, prompts related to embroidery and folk-art practices often turned distinct regional aesthetics into generalized craft imagery. One participant generated image and video outputs based on her embroidery practice, allowing local histories and material traditions to inform the AI-generated outputs.

Participants also created variants of the Lascaux Cave Paintings, a style intentionally introduced to examine how generative AI systems engage with forms of cultural heritage that fall outside dominant contemporary or commercially visible aesthetic categories.

While baseline AI systems often generated generic prehistoric cave scenes, participants were able to create outputs that better reflected the textures, animal forms, and visual style of the original paintings.

One participant reflected that “the conversation wasn’t only on developing or improving AI models but on questions of bias and representation embedded within the dataset itself,” highlighting how workshops became spaces for critical reflection on the cultural assumptions shaping generative AI systems. These interactions prompted broader discussions concerning how generative systems encode dominant cultural assumptions and flatten contextual forms of knowledge [3,4].

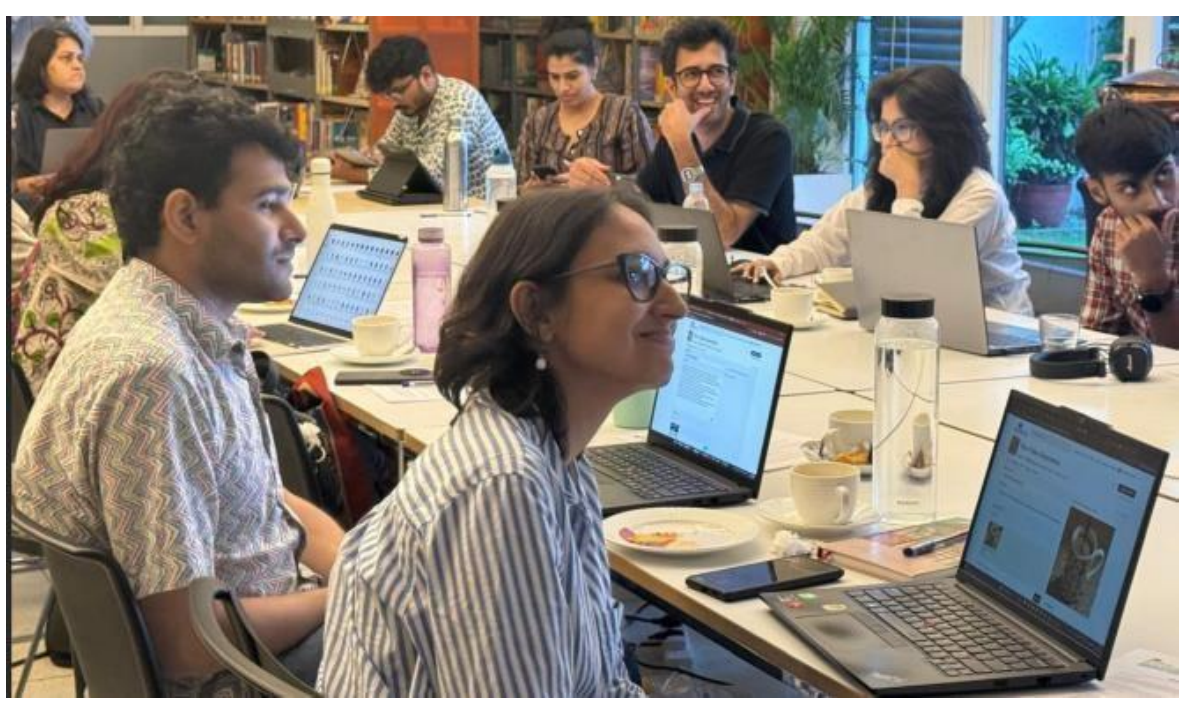

Figure 2: Participants During the Beyond Bias Promptathon in New Delhi, October 2025

Participants generated alternative outputs reflecting more situated cultural interpretations like reinterpretations of Mughal miniature traditions depicting women protesting in New Delhi, riding bicycles, and engaging contemporary political contexts.

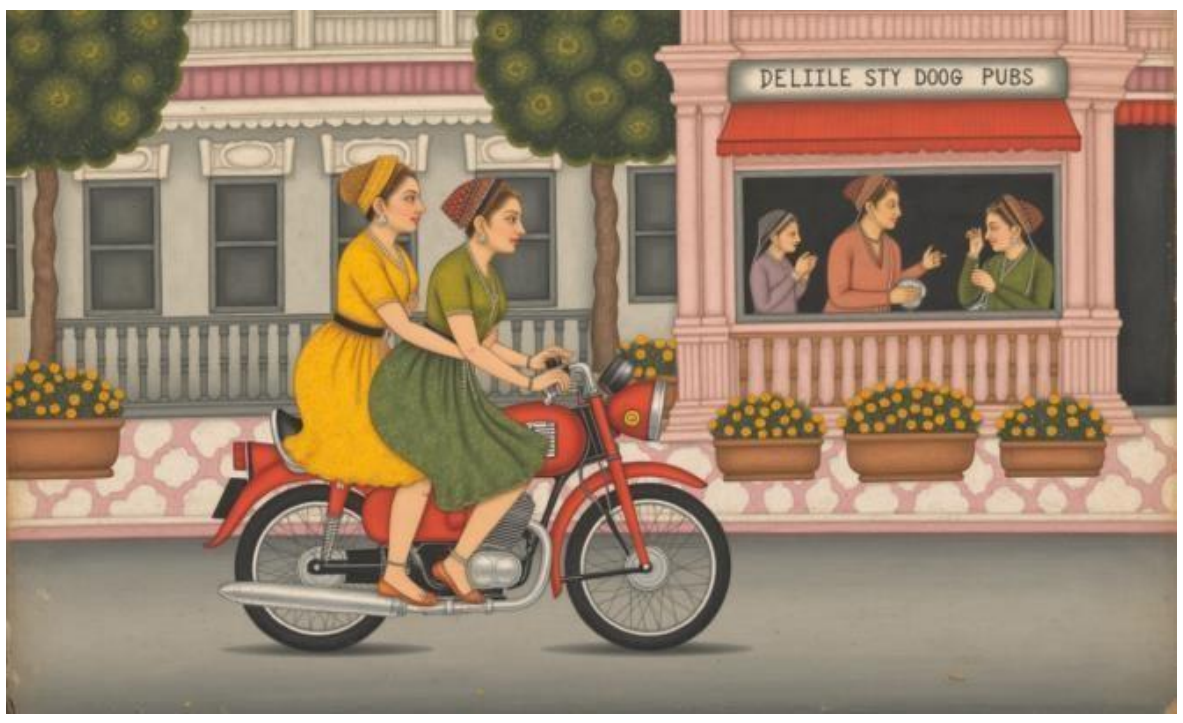

Figure 3: Women Riding a Bike - Output in Mughal Miniature Style

Across workshops, participants conducted more than 1,600 generation runs involving iterative experimentation with prompts, datasets, and fine-tuned models. Workshops functioned as iterative cycles of co-design in which participant feedback directly informed subsequent tooling decisions, dataset practices, and governance discussions [5]. Participants increasingly described the tools not simply as systems for image generation, but as creative scaffolds for reflecting on authorship, representation, and cultural interpretation itself.

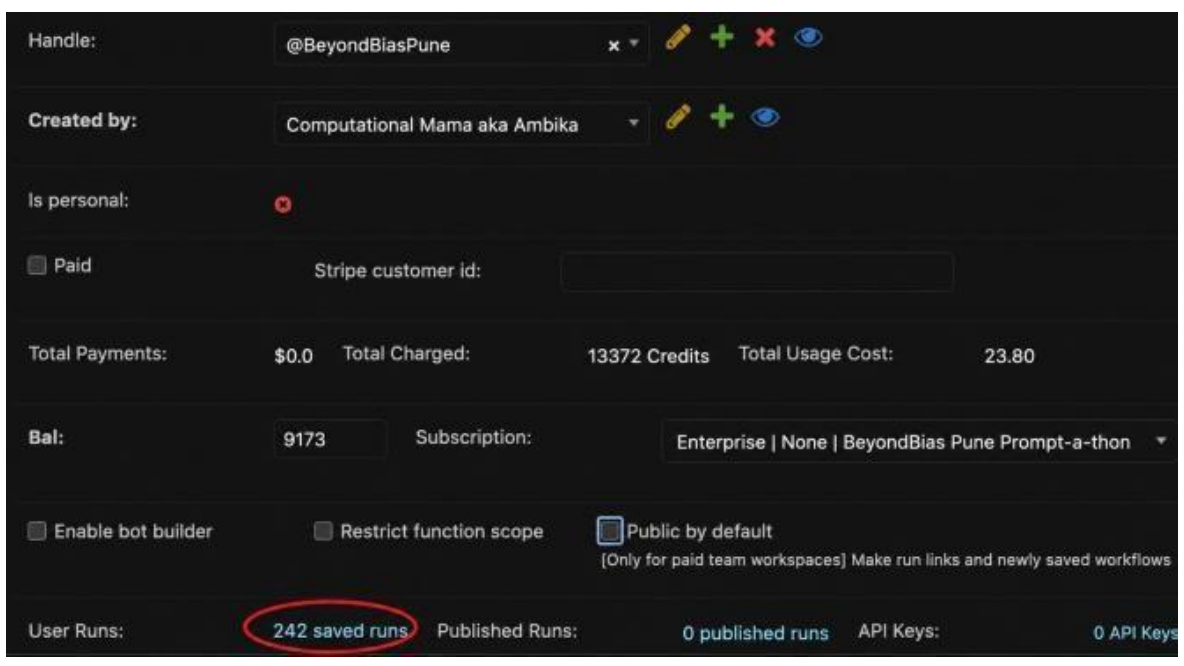

Figure 4: Metrics from the Beyond Bias Workshop in Pune with 242 runs

## 4 A Framework for Cultural AI

Beyond Bias emphasized the importance of contextual understanding, interpretive agency, and community stewardship within generative systems. Based on workshop reflections, participatory tooling practices, and collaborative experimentation, we propose four interconnected dimensions for cultural AI:

- **Cultural Integrity:** Cultural AI systems should preserve contextual meaning, symbolism, provenance, and historical relationships embedded within cultural forms rather than reducing culture to surface-level visual aesthetics [4]. Participants repeatedly emphasized that generated outputs may appear stylistically accurate while still failing to represent the social, political, and historical meanings associated with cultural practices.
- **Participation and Agency**: Participants emphasized that communities should be able to shape AI systems directly through dataset curation, fine-tuning, prompting workflows, and governance discussions. Participatory approaches therefore position cultural practitioners not only as users of AI systems, but as co-creators of generative infrastructures [5].
- **Interpretive Capacity**: Participants frequently used generated outputs as prompts for reflection and reinterpretation rather than as final creative artefacts. Workshops became spaces where participants critically examined how prompts, datasets, and interfaces shaped representation itself. Cultural AI systems therefore require the capacity to support dialogue and critical engagement rather than simply automating cultural reproduction [4].

- **Stewardship and Governance**: Participants also foregrounded concerns surrounding ownership, consent, environmental impact, and cultural extraction within generative AI systems [1,6].

## 5 Conclusion

Participatory approaches to cultural AI create new possibilities for representation and creative agency while also introducing tensions concerning ownership, labor, consent, and scalability. A participant described the initiative as “wielding a machete through the jungle of AI bias with fellow explorers,” emphasizing the experimental, collaborative, and uncertain nature of navigating AI systems.
Although collaborative fine-tuning enabled more culturally situated outputs, participants frequently reflected on the limitations of working within commercial foundation-model ecosystems trained on opaque datasets [1,6].

Beyond Bias suggests that participatory cultural AI requires moving beyond harm mitigation toward systems that support cultural integrity, interpretive agency, and community stewardship. Rather than treating culture as extractable data, cultural AI frameworks that foreground reflection may enable communities to become active co-creators of AI.

## ACKNOWLEDGMENTS

We thank Dev Aggarwal, CTO of Gooey.AI and Computational Mama (Head of Developer Relations) for their technical guidance and support throughout the Beyond Bias initiative. We also gratefully acknowledge Goethe-Institut / Max Mueller Bhavan New Delhi for its partnership and collaboration in enabling the Beyond Bias workshops and participatory research that informed this work.